# Rapid appearance of domains upon phase change in $KNbO_3$ - a TEM in-situ heating study.

A.F. Mark, W. Sigle

Stuttgart Center for Electron Microscopy, Max Planck Institute for Intelligent Systems, Heisenbergstrasse 3, 70569 Stuttgart, Germany

**Abstract**

TEM specimens from potassium niobate single crystals were observed while being heated in a TEM. DWs and dislocations were observed; the DWs were mobile. In certain cases the DWs became pinned by the dislocations, at least for a short time, most likely due to interaction of strain fields. Both phase changes were observed with accompanying rapid appearance of new domain patterns.

## 1. Introduction

Potassium niobate ($KNbO_3$ or KNO) is a potentially useful ferroelectric perovskite oxide for replacing lead-based perovskites in a variety of applications. As part of a larger study on the mechanical properties of KNO [1] and their interaction with the electrical response, TEM in-situ heating experiments were performed.

The aim was to observe interactions between moving domain walls and dislocations, and this was accomplished. The two phase transformations undergone by KNO were also observed, with interesting results.

## 2. Material

Potassium niobate is a ferroelectric ceramic material of the perovskite type. Above 435°C [2] it has a cubic structure and is not ferroelectric. Between 435°C and room temperature it undergoes two phase transformations, to a tetragonal structure and then to an orthorhombic structure. At low temperatures (below -10°C) it has a rhombohedral structure.

The different structures arise due to displacements of the oxygen octahedra and the Nb atoms. These relative displacements give rise to the polarity, the ferroelectric behaviour. As summarized in Table 1 in the tetragonal structure the offset is in the [001] direction, in orthorhombic it is in the [011] direction and in rhombohedral it is in the [111] direction [3]. Details of the phase changes are given in Table 1.

In this and similar materials it is conventional to refer to directions based on the cubic system. For the tetragonal structure this is straight-forward; for the orthorhombic structure the pseudocubic structure is oriented with $[100]_o$ par to $[100]_{pc}$, $[011]_o$ to $[010]_{pc}$ and $[0\text{-}11]_o$ to $[001]_{pc}$. The pc subscript is to be assumed in this paper, unless otherwise is specified.

Table 1. Phase changes in KNbO$_3$

| Phase | rhombohedral | orthorhombic | tetragonal | cubic |
|---|---|---|---|---|
| Transformation temperature | -10°C | 225°C | 435°C | |
| Polarization direction | [111] | [011] | [001] | - |
| Unit cell volume | 6.48 x 10$^{-3}$ nm$^3$ | 6.47 x 10$^{-3}$ nm$^3$ | 6.49 x 10$^{-3}$ nm$^3$ | 6.52 x 10$^{-3}$ nm$^3$ |

**3. Method**

A single crystal of KNO was purchased from FEE GmbH. Two TEM specimens were prepared by diamond wire slicing, grinding and polishing, followed by ion milling, with liquid nitrogen cooling. The specimens were cut normal to the [100] direction.

The specimens were observed in a JEOL 4000FX TEM equipped with a heating holder, and operated at 400kV. The specimens were heated in stages to approximately 465°C and continuously observed. The image was recorded digitally. The temperature was monitored by a Type R thermocouple spot welded to the specimen holder and displayed on the heating holder control unit. It was recorded by hand on the first occasion and by video camera on the second.

**4. Results and Discussion**

The density of dislocations in the specimens was fairly low but in both cases both dislocations and DWs were observed. Under the influence of the electron beam and the heating and cooling the dislocations remained immobile, for the most part, while the DWs moved, appeared, and disappeared.

<u>4.1 Interaction of domain walls and dislocations</u>

Two interesting effects were observed. The first was that domain boundaries had inconsistent interaction with dislocations. In most cases, domain walls moved across dislocations smoothly. In some cases there may have been a slight hesitation in the movement of a domain wall as it contacted a parallel dislocation and then moved away from it. There were other cases where a dislocation appeared to cause splitting of a domain, or reorientation of a domain wall.

Example videos are:

heat_198-to-224.mp4

heat_170-to-206.mp4

(See also supplementary video: heat_419-to-424.mp4)

Figures 1 and 2 show still images captured before and after the relevant movements.

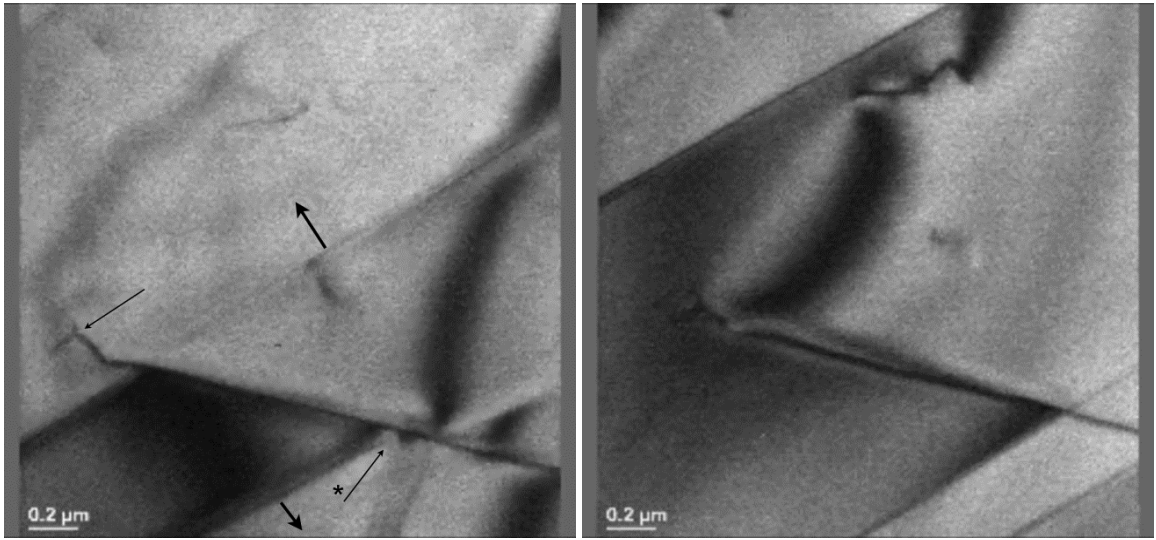

          (a)                                          (b)

**Figure 1**. Still images captured during heating before, at about 200°C (a), and after, at about 217°C (b), the upper DW moved, with a pause, over the parallel dislocation segment, marked with the upper thin arrow in (a). The thicker arrows in (a) indicate the directions of DW movement. The thin arrow marked with * points to a split in the domain at the dislocation. Note its absence in (b).

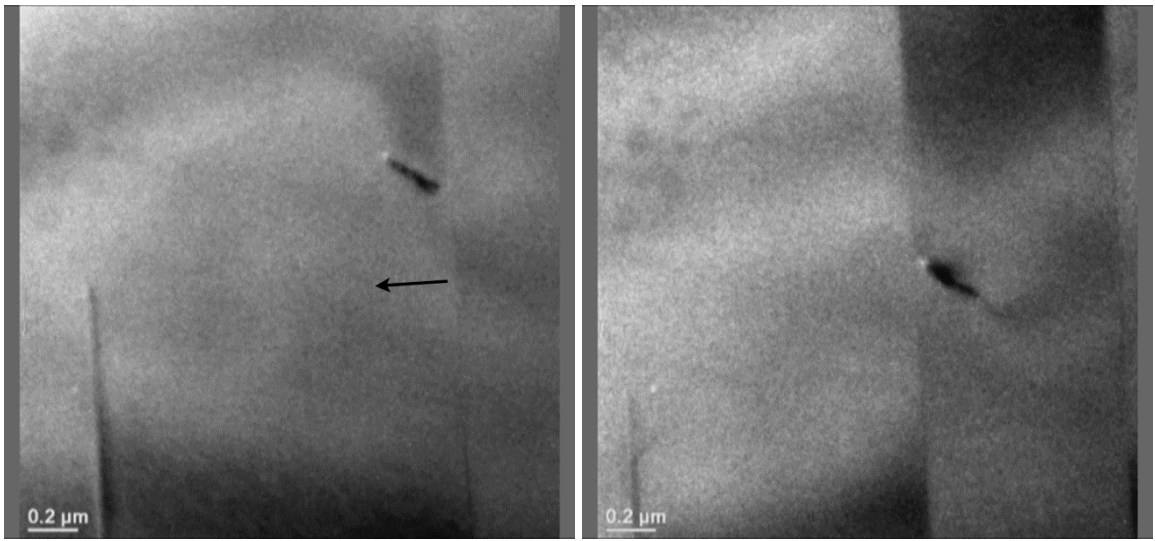

          (a)                                          (b)

**Figure 2**. Still images captured during heating from about 170°C to 205°C before (a) and after (b) the DW moved smoothly over the dislocation. Arrow in (a) indicates the direction of DW movement.

The simplest explanation for the interaction of some DW-dislocation pairs and not others is the different orientations of relevant stress or strain fields. In cases where the DW strain field relieves some of the strain around the dislocation the DW may be pinned by the dislocation [4].

In general, for a straight segment of a dislocation there is a non-uniform stress (or strain) field around the dislocation. There may be compressive stress on one side and tensile on the other (edge dislocation)[5] or shear stresses differing in sign (screw dislocation)[5,6]. This non-symmetric strain field may lead to the strain energy of the dislocation being most effectively lowered by twinned domains on either side with the DW at the dislocation.

Previous researchers have discussed the nucleation of domain walls near dislocations in $KNbO_3$ under an applied electric field. They note that the new domain is favourable due to the strain energy of the dislocation being lowered by the particular orientation of the new polar axis, and thus of the distortion of the unit cell [7]. Other researchers studying $BiFeO_3$ observed particular DW types pinned at dislocations with particular Burgers vectors, while other similar dislocations had no pinned DWs [8]. Lubk *et al* do mention that their results are, of course, not directly transferable to other materials with only similar structures. There is none-the-less reasonable support from both studies mentioned for the idea that certain combinations of dislocations and DWs will interact on the basis of strain relief while others will not.

A second possible reason for the pinning of a DW by a dislocation is local polarization in the dislocation core. Early molecular statics simulations by Hirel [9] indicate that the polarization inside the dislocation core is different to that in the surrounding bulk. In his models, in the case of an edge dislocation with a 180° difference in polarization direction between the core and the bulk, this resulted in the nucleation of a new domain when an electric field was applied. With a similar screw dislocation a new domain did not nucleate. However, this dislocation did act as a barrier to DW movement.

4.2 Phase changes

The second interesting phenomenon that was observed during the in-situ experiments was the rapid appearance of new domains upon phase change, particularly upon cooling. From 460°C, the specimen was cooled past the higher transformation temperature and at about 420°C new domains suddenly appeared. Their growth was so rapid as to be indistinguishable in the video. The specimen was cooled further, past the lower transformation temperature and at about 185°C the existing domain pattern was replaced. New domains appeared in the image and rapidly grew to fill the entire imaged area. In this second case the growth was rapid but observable.

Example videos are:

cool_428-to-417.mp4

cool_197-to-183.mp4

Figures 3 and 4 show still images captured before and after the transformations.

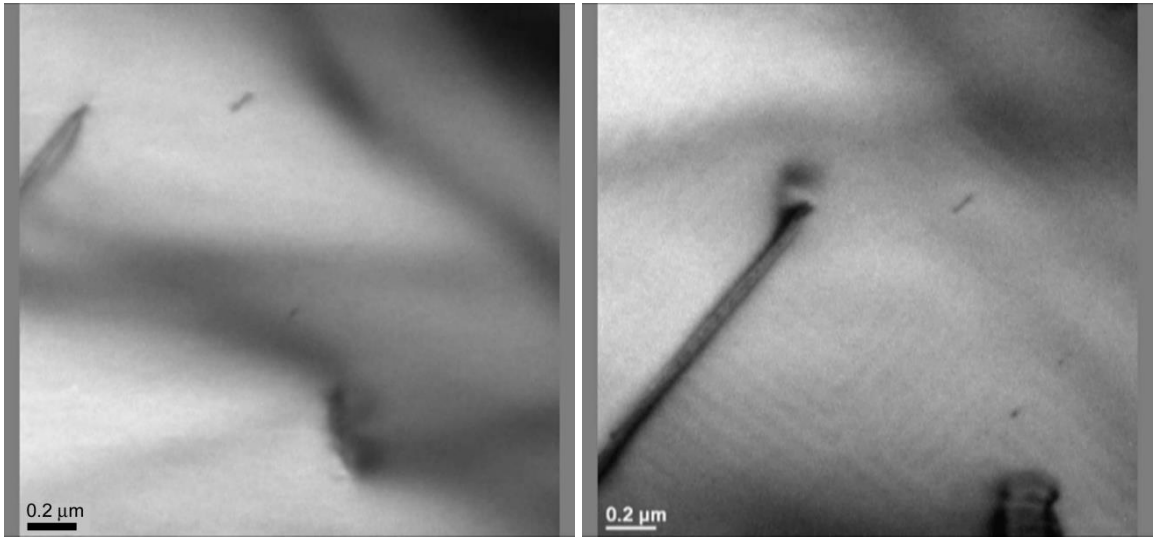

**Figure 3**. Still images captured during cooling before (a) and after (b) the phase transformation and sudden appearance of domains at about 420°C.

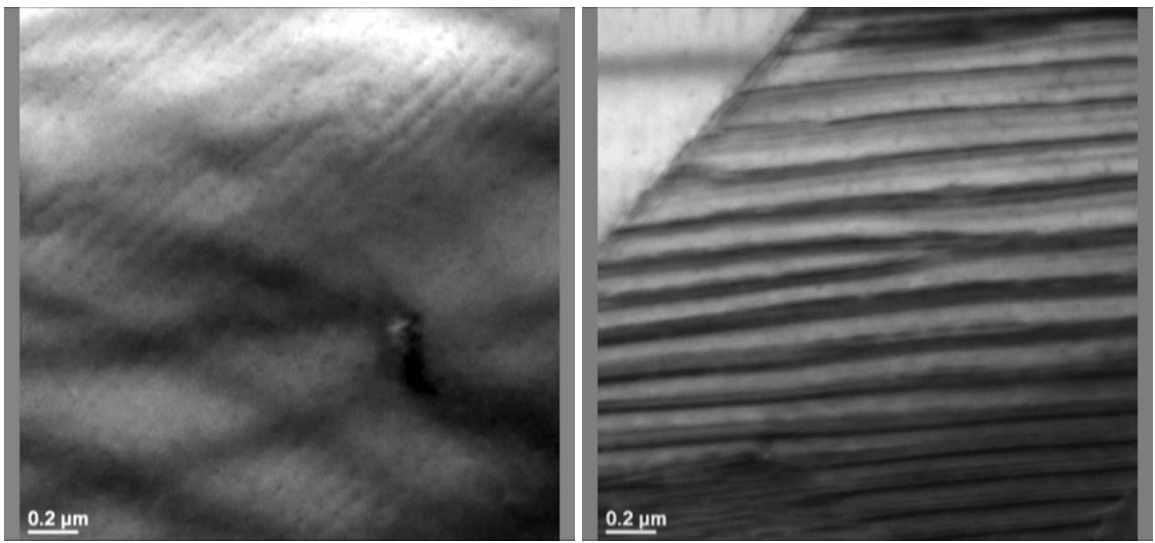

**Figure 4**. Still images captured on cooling before (a) and after (b) the sudden change in domain pattern at about 185°C.

The observations here match some of the few other examples in the literature. The fine domain pattern visible after the cubic-tetragonal phase change is also reported by Li *et al* [10]. Popoola and Kriven report the instantaneous appearance of a similar fine domain pattern above the temperature of the phase change, but in their specimen a second domain pattern then formed by nucleation and growth starting at a temperature a few degrees lower [11]. They report that the lower phase transformation occurred with a front that passed through the specimen instantaneously, once initiated. A specific orientation relationship is described in their paper but no images are presented of the domain pattern after transformation [11].

## 5. Conclusions

Phase transformations and DW movement were successfully observed in KNbO$_3$ TEM specimens. Under moderate heating rates at low strains the DWs were mobile, while the dislocations were not. DWs and dislocations with some particular orientation relationships appeared to interact. The DWs were momentarily, at least, pinned by the dislocations, most likely due to the interaction of strain fields.

Phase transformations occurred rapidly and resulted in both cases in the sudden appearance of new domain patterns in the new phase.

For the purposes of using these results to help understand the mechanical properties of KNbO3 several points can be mentioned. At elevated temperature and low strains the response to strain is DW movement. The appearance of a dense pattern of DWs upon phase transformation will contribute to greater internal strain as the material is cooled and the anisotropic thermal expansion (contraction) occurs in the different domains.

From the perspective of a user wishing to control and switch the ferroelectric polarity the interaction of the DWs with the dislocations deserves further study. Models are being developed for this purpose [12].